\documentstyle [12pt]{article}

\title{Non trivial critical exponents for finite temperature chiral 
transitions at fixed total fermion number}

\author{F. S. Nogueira \thanks{email: nogueira@cpht.polytechnique.fr} \\
 Centre de Physique Th{\' e}orique, Ecole Polytechnique \\ 
 F91128 Palaiseau Cedex, FRANCE \\ and \\ 
        M. B. Silva Neto \thanks{e-mail: sneto@lafex.cbpf.br} and  
        N. F. Svaiter \thanks{e-mail: nfuxsvai@lafex.cbpf.br} \\
 Centro Brasileiro de Pesquisas F\'{\i}sicas - CBPF\\ Rua Dr.Xavier
 Sigaud 150, Rio de Janeiro - RJ, CEP: 22290-180, BRAZIL}

\begin{document}

\maketitle

\newpage


\begin{abstract}

We analyze the finite temperature chiral restoration transition of the 
$(D=d+1)$-dimensional Gross-Neveu model for the case of a large
number of flavors and fixed total fermion number. 
This leads to the study of the model with a nonzero imaginary chemical 
potential. In this formulation of the theory, we have obtained that, in the 
transition region, the model is described by a chiral conformal field theory 
where the concepts of dimensional reduction and universality do apply due
to a transmutation of statistics which makes fermions act as if they were
bosons, having zero energy. This result should be generic for theories 
with dynamical symmetry breaking, such as Quantum Chromodynamics.

\end{abstract} 

\vfill

Key-words: Dynamical symmetry breaking; Finite temperature QFT;
Dimensional reduction; Phase transitions; Chemical potential.

Pacs numbers: 11.10.Ef, 11.10.Gh 

\newpage\baselineskip .37in


\section{Introduction}

Recently, Koci\'c and Kogut \cite{Kocic-Kogut} have shown that, in the limit 
of a large number of flavors $(N)$, the universality class of the finite
temperature chiral symmetry restoration transition in the $3D$ Gross-Neveu
model is mean field theory, in contrast to the "standard" sigma model 
scenario which predicts the $2D$ Ising model universality class. 
The responsible for the breakdown of the "standard" scenario, dimensional
reduction plus universality \cite{Pisarski-Wilczek-Rajagopal}, would be
the absence of canonical scalar fields in the model, since the $\sigma$
auxiliary field has a composite nature. Indeed, as explained
in \cite{Kocic-Kogut}, both the density and the size of
the $\sigma$ meson increase with the temperature in such a way that close
to the restoration temperature the system is densely populated with 
overlapping composites. In opposition to what happens in the dynamics of
BCS superconductors, here the constituent fermions are essential degrees of
freedom, even in the scaling region. Thus, we conclude that, for this model,
the effect of compactifying the Euclidean time direction is simply to
regulate the infrared behavior and supress fluctuations. This fermionic model 
was first analyzed in \cite{Rosenstein-Warr-Park-1} and it was further
argued that the results would not be an artifact of the large $N$ limit 
\cite{Rosenstein-Speliotopoulos-Yu}. (In addition, it was shown how the 
Ising point is recovered in 4D Yukawa models beyond the leading order in
the $1/N$ expansion and why this should not happen in Gross-Neveu models.)
Lattice simulations of this 3D model have verified the 
predictions of the large $N$ expansion at zero temperature, at nonzero
temperature, and at nonzero chemical potential separatedly 
\cite{Hands-Kocic-Kogut}. The results for the critical indices have 
been checked and improved by larger scale simulations enhanced by histogram 
methods \cite{Karkkainen-Lacaze-Lacock-Peterson}. The conclusion was that 
the data are in excellent agreement with mean field theory and rule
out the Ising model values. However, the question concerning the nature of the 
transition for large but finite $N$ in Gross-Neveu models remained open 
until the recent work of Reisz \cite{Reisz} where he suggests
that the mean field exponents are indeed an artifact of the large 
$N$ limit. By a combination of different techniques, 
like $1/N$ expansion, dimensional reduction and high-temperature series, 
it is shown how the 2D Ising model exponents are recovered beyond the 
leading order in the $1/N$ expansion in such a way that the "standard" 
scenario is still valid, if $N$ is not large enough. 

In this letter we shall study further the large $N$ limit of the Gross-Neveu 
model. We will show that, if we fix the particle number, it is possible to 
get non-trivial exponents even if $N$ is infinitely large. This third set
of exponents is consistent with the "standard" scenario with the difference
that, as we will see, the reduced theory is not the linear sigma model but,
instead, again a four fermi theory for massless fermions. According to the 
"standard" linear sigma model scenario, if dimensional reduction plus 
universality arguments hold, the finite temperature transition of the 
$D$-dimensional Gross-Neveu model would lie in the same
universality class of the $(d=D-1)$-dimensional linear sigma model.
This is not what is found and the reason why is easy
to understand in the finite temperature Matsubara formalism: fermions do not
have a zero frequency Matsubara mode. (The introduction of a real chemical
potential would not change this results since it would simply cause a
shift of the frequencies in the imaginary axis.) On the other hand, we will
show that the imposition of a constraint which fixes the total fermion
number gives rise to a fermionic zero frequency Matsubara mode. In this 
case, the transition has neither mean field nor scalar critical indices
but rather chiral indices, a feature of theories with long range interactions.
The basic idea is to introduce a pure imaginary chemical
potential in the partition function of the model. The imaginary chemical 
potential arises when we impose a constraint which fixes the particle number
\cite{Weiss}. 
This is generally done by inserting in the functional integral a functional 
delta function which inforces exactly the constraint. As a consequence, an 
auxiliary field is introduced in the problem. Usually, saddle-point solutions 
for this auxiliary field are imaginary implying in this way a real chemical 
potential. (This happens for example in the non-linear sigma model.) However, 
as we will see, real solutions are also allowed in the Gross-Neveu 
model with constrained particle number. Such a solution can be thought as 
an imaginary chemical potential.
In this new "physical" system, we do have a 
zero frequency Matsubara mode associated to fermions affecting the infrared
sensitiveness of thermodynamic quantities and, consequently, changing its
critical indices. We observed that the new critical exponents are exactly
the same as those corresponding to a zero temperature dimensionally reduced
effective theory. In other words, after the decoupling of nonzero frequency Matsubara modes (dimensional reduction) the effective theory for the light 
degrees of freedom has a set of critical indices identical to that of the 
original zero temperature one (universality), where one simply has to replace 
$D=d+1$ by $d$.


\section{The Gross-Neveu model at fixed total fermion number}

We start by considering a $(D=d+1)$-dimensional fermionic model 
with fixed total fermion number $B$ described by the canonical partition 
function
\begin{equation}
Z= Tr \left\{ e^{-\beta H} \delta(\hat{N}-B) 
          \right\},
\label{part-func} 
\end{equation}
where $\beta$ is the inverse of the temperature, $H$ is some Hermitian 
Hamiltonian and
\begin{equation}
\hat{N}=\int d^{d}{\bf x} (\psi^{\dag} \psi)
\end{equation}
is the fermion number operator. We can rewrite the partition function as a 
functional integral over Grassmann fields
\begin{eqnarray}
Z&=&\int_{B.C.} D\psi^{\dag}D\psi {\,} 
\delta\left(\int d^d {\bf x}(\psi^{\dag}\psi)-B\right)
e^{-A(\psi^{\dag},\psi)}\nonumber\\
&=&\int_{B.C.} D\psi^{\dag}D\psi D\theta {\,} 
e^{-\tilde{A}(\psi^{\dag},\psi,\theta)},
\label{functional-integral}
\end{eqnarray}
where we have used the integral representation of the functional delta 
function. In the above equation $A$ is the action of the unconstrained 
model while $\tilde{A}$ is given by
\begin{equation}
\tilde{A}=A+\int d^d {\bf x}\int_{0}^{\beta}d\tau{\,}\overline{\psi}
i\frac{\theta}{\beta}\gamma_{0}\psi+
\frac{iB}{\beta}\int_{0}^{\beta}d\tau{\,}\theta.
\end{equation}
where the Euclidean gamma matrices are antihermitian and satisfy 
$\{\gamma_{\mu},\gamma_{\nu}\}=-2\delta_{\mu\nu}$.
As usual, the Grassmann fields are integrated with antiperiodic boundary 
conditions (B.C.) with respect to the Matsubara time. Note that the Bose field 
$\theta$ is a function of $\tau$ but not of ${\bf x}$. This is expected since 
$\theta$ is an auxiliary field related to a global constraint with respect to 
space rather than a local one. 

We will use as fermionic action, the 
action for the massive Gross-Neveu model \cite{Gross-Neveu} with $N$ 
components. By rewriting it in its standard auxiliary field version we get the 
following Euclidean action for the fixed fermion number model
\begin{eqnarray}
\label{action}
S_{E}(\psi,\overline{\psi},\sigma,\theta)&=&\int_{0}^{\beta}d\tau
\int d^{d}{\bf x}
\left[\overline{\psi}\left(-\gamma^{0}\frac{\partial}{\partial\tau}+
i\vec{\gamma}{\cdot}\vec{\nabla}+m+g\sigma+
i\frac{\theta}{\beta}\gamma^{0}\right)\psi
\right.
\nonumber\\
&+&\left.\frac{1}{2}\sigma^2+\delta^d({\bf x})\frac{iB}{\beta}\theta\right].
\end{eqnarray}
For bare fermion mass $m=0$, there is a discrete chiral symmetry
$\psi \rightarrow \gamma_{s}\psi$, 
$\bar{\psi} \rightarrow -\bar{\psi}\gamma_{s}$ which is spontaneously
broken whenever a non-vanishing condensate $<\bar{\psi}\psi>$ is generated.
The condensate
\begin{equation}
\left< \bar{\psi}(x)\psi(x)\right> \propto
\int {\cal D}{\theta}{\cal D}{\sigma}{\cal D}{\bar{\psi}}{\cal D}{\psi}
\left[ \bar{\psi}(x)\psi(x) \right]{\,}
e^{-S_{E}(\psi,\psi^{\dag},\sigma,\theta)},
\end{equation}
serves as an order parameter of the transition. It is well known that
this is the role played by the auxiliary field $\sigma$ in eq. (\ref{action}). 
Thus, we
will study the phase structure of the chiral symmetry restoration transition
with the aid of the order parameter $\sigma$. Note that eq. (\ref{action}) is 
invariant by a partially local gauge transformation. By this we mean  a 
gauge transformation which is local in the Matsubara time and global in 
the space. In fact, the transformation $\psi\to e^{i\phi}\psi$, 
$\overline{\psi}\to e^{-i\phi}\overline\psi$, leaves the action invariant 
provided that $\theta\to\theta+\beta\partial\phi/\partial\tau$. Thus, 
$\theta$ plays the role of a "gauge field". It should be observed that, due 
to the periodicity of the $\theta$ with respect to the Matsubara time, the 
last term in eq. (\ref{action}) is also gauge invariant: 
$\int_{0}^{\beta}d\tau\partial\phi/\partial\tau=\phi(\beta)-\phi(0)=0$. 
Later we shall discuss the role of this gauge symmetry with respect to 
the zero mode of the fermion field.


\section{Saddle-point evaluation of the functional integral}

Integrating out exactly the fermions in eq. (\ref{functional-integral})
we obtain the effective action 
\begin{eqnarray}
\label{action1}
S_{eff}&=&-NTr\log\left(-\gamma^{0}\frac{\partial}{\partial\tau}+
i\vec{\gamma}{\cdot}\vec{\nabla}+m+g\sigma+
i\frac{\theta}{\beta}\gamma^{0}\right)
\nonumber\\
&+&\int_{0}^{\beta}d\tau\left(\int d^{d}{\bf x}{\,}
\frac{1}{2}\sigma^2+\frac{iB}{\beta}\theta\right).
\end{eqnarray}
Now we perform a saddle-point evaluation of the functional integral. From 
the structure of $S_{eff}$ we know that the saddle-point solution is  
exact at large $N$. If $N$ is large enough, the fluctuations of the $\sigma$ 
and $\theta$ auxiliary fields average out and for this reason we look for 
uniform saddle-point solutions. By imposing stationarity with respect to 
$\sigma$ we obtain the gap equation
\begin{equation}
\label{sp1}
\Sigma=m+\frac{4g^{2}}{\beta}
       \sum_{n=-\infty}^{\infty}
       \int^{\Lambda}\frac{d^{d}{\bf q}}{(2\pi)^{d}}
       \frac{\Sigma}{{\bf q}^{2}+(\rho-\bar{\omega}_{n})^{2}+\Sigma^{2}},
\end{equation}
where $\bar{\omega}_{n}=\frac{2\pi}{\beta}(n+\frac{1}{2})$, 
and we have used the fact that,
to the leading order, the fermion self-energy $\Sigma$ comes from the 
$\sigma$ auxiliary field tadpole: $\Sigma=m-g^{2}<\bar{\psi}\psi>$. Here
$\Lambda$ is an ultraviolet cutoff and we have defined $\rho=\theta/\beta$.  
To find the temperature dependence 
of the order parameter $\sigma$ one has to solve the gap equation near 
criticality. The critical temperature is determined by
\begin{equation}
\label{Tc}
1=\frac{4g^{2}}{\beta_{c}}
       \sum_{n=-\infty}^{\infty}
       \int^{\Lambda}\frac{d^{d}{\bf q}}{(2\pi)^{d}}
       \frac{1}{{\bf q}^{2}+(\rho_{c}-\bar{\omega}_{nc})^{2}},
\end{equation}
with $\bar{\omega}_{nc}=\frac{2\pi}{\beta_{c}}(n+\frac{1}{2})$. 
On the other hand, stationarity with respect to $\theta$ gives us

\begin{equation}
\label{B}
\frac{N}{\beta}Tr[\gamma^{0}\cdot S_{F}(\tau-\tau';{\bf x}-{\bf x}')]=B, 
\end{equation}
where $S_{F}$ denotes the fermion propagator given by the inverse of the 
argument of the $\log$ in Eq.(7). Eq.(\ref{B}) is equivalent to the 
statement $<\psi^{\dag}\psi>=B$ which fixes the mean number of fermions. Note 
that some care is needed to evaluate the trace in (\ref{B}) since 
$\gamma^{0}\cdot S_{F}$ is divergent for $\tau=\tau'$ (for 
${\bf x}={\bf x}'$ there is no problem due to the presence of the 
ultraviolet cutoff $\Lambda$). 
The same kind of difficulty is encountered in the theory of fermion systems, 
for instance Fermi liquid theory, when defining expectation values of 
particle number operators using the one-particle propagator 
\cite{Negele-Orland}. In order to remedy this problem it is sufficient 
to evaluate $\gamma^0\cdot S_{F}(\tau-\tau';0)$ with a temporal 
point-splitting  
$\tau-\tau'=\eta$ where $\eta$ is a small positive number. 
After we let $\eta\to 0+$ getting a finite result. This means that if we 
write the Fourier representation of $\gamma^0\cdot S_{F}(\tau-\tau';0)$, the 
limit $\tau-\tau'\to 0$ does not commute with the Matsubara sum.  
This is the standard procedure used in many-particle 
systems to evaluate expectation values of one-body operators 
\cite{Negele-Orland}. Evaluating explicitly the trace in (\ref{B}), we obtain    
\begin{equation}
\label{sp2}
\frac{1}{\beta}\sum_{n=-\infty}^{\infty}\int^
{\Lambda}\frac{d^{d}{\bf q}}{(2\pi)^{d}}
\frac{e^{i\bar{\omega}_{n}\eta}(\rho-\bar{\omega}_{n})}
     {{\bf q}^2+(\rho-\bar{\omega}_{n})^2+\Sigma^2}=-\frac{ib}{4},
\end{equation}
where $\eta\to 0_{+}$ and we have introduced the barion density 
$b=B/(N\Omega)$, $\Omega$ being the volume. 
Note that the sum is divergent without the convergence 
factor $e^{i\omega_{n}\eta}$ 
even if $\rho=0$. It is divergent simply because it is not defined (like 
the series $1-1+1-1+...$) \cite{discussion}.   

The Matsubara sums in eqs.(\ref{sp1}) and (\ref{sp2}) are done by contour 
integration and we obtain
\begin{equation}
\label{sp11}
\Sigma=m+2g^2\Sigma\int^{\Lambda}
\frac{d^{d}{\bf q}}{(2\pi)^{d}}\frac{1}{\omega}
\left(\frac{1}{e^{i\theta}e^{\beta\omega}+1}-
\frac{1}{e^{i\theta}e^{-\beta\omega}+1}\right),
\end{equation}
for eq. (\ref{sp1}) and
\begin{equation}
\label{sp21}
2\int^{\Lambda}\frac{d^{d}{\bf q}}{(2\pi)^{d}}
\left(\frac{1}{e^{i\theta}e^{\beta\omega}+1}
+\frac{1}{e^{i\theta}e^{-\beta\omega}+1}\right)=b,
\end{equation}
for eq. (\ref{sp2}) and we have defined $\omega=\sqrt{{\bf q}^2+\Sigma^2}$.
One saddle-point solution for $\theta$ is $\theta=i\beta\mu$, $\mu$ 
playing the role of an ordinary chemical potential. This solution is 
non-singular in the infrared since it does not generate any zero mode. In 
this situation the exponents are mean field. However, the solution 
$\theta=(2m+1)\pi$, where $m$ is an integer, is also possible if 
$B$ is even. (We need $B$ even in order to ensure that the free energy is 
a real quantity.) Note that every solution of this form corresponds to the 
same free energy and therefore we can choose any of them, say $\theta=\pi$. 
This particular solution is related to any other solution 
$\theta=(2m+1)\pi$ by a gauge transformation. The existence of such 
solutions is a consequence of the partially local gauge symmetry we have 
discussed in the beginning of this paper.


\section{Scaling properties at fixed fermion number}

Let us discuss the implications of the solution $\theta=\pi$ to
the critical behavior of the model. First, by using eq. (\ref{Tc}) together 
with eq. (\ref{sp1}) we can rewrite the gap equation for the order parameter 
$\sigma$ as
\begin{equation}
\frac{m}{\Sigma}+\frac{4g^{2}}{\beta_{c}}
       \sum_{n=-\infty}^{\infty}
       \int^{\Lambda}\frac{d^{d}{\bf q}}{(2\pi)^{d}}
       \frac{\left({\bf q}^{2}+(\rho_{c}-\bar{\omega}_{nc})^{2}\right)
             \left( \frac{\beta_{c}}{\beta} \right) -
             \left({\bf q}^{2}+(\rho-\bar{\omega}_{n})^{2}+\Sigma^{2}\right)}
       {\left({\bf q}^{2}+(\rho_{c}-\bar{\omega}_{nc})^{2}\right)
        \left({\bf q}^{2}+(\rho-\bar{\omega}_{n})^{2}+\Sigma^{2}\right)}
       =0.
\label{gap-equation-1}
\end{equation} 
This form of the gap equation is particularly well suited for extracting
critical indices since the problem reduces to the power counting of the 
infrared divergences in the integral over ${\bf q}$ \cite{Zinn-Justin}.
For $\theta=\pi$ we see that the above integral is infrared divergent as 
$\Sigma\to 0$. The critical indices are defined by
$\left. \left< \bar{\psi}\psi \right> \right|_{m \rightarrow 0}
\sim t^{\beta}$, 
$\left. \left< \bar{\psi}\psi \right> \right|_{t \rightarrow 0}
\sim m^{1/\delta}$,
$\left. \left< \Sigma \right> \right|_{m \rightarrow 0}
\sim t^{\nu}$ etc and, since 
$\Sigma\sim<\bar{\psi}\psi>$, $\beta=\nu$ to the leading order. 
Above four dimensions the
integral is infrared finite and the scaling is mean field. Below four 
dimensions, however, the $\Sigma \rightarrow 0$ limit is singular and
the integral scales as
$\left( \bar{\omega}_{nc}^{2}t^{2}+\Sigma^{2} \right)^{\frac{d-2}{2}}$. 
At a critical point $t=0$ away from the chiral limit we
have $m\sim\Sigma^{d-1}$. Thus, we can easily see that, below four 
dimensions the $\beta$ and $\delta$ exponents are
\begin{equation}
\beta=\frac{1}{d-2}, {\;}\delta=d-1.
\label{exponents}
\end{equation}
The remaining exponents are easily obtained as well, 
$\eta=4-d$ and $\gamma=1$, and one 
can check that they obey hyperscaling.
If we compare these indices with those obtained from the zero temperature
case: $\beta=\frac{1}{D-2}, \delta=D-1, \eta=4-D, \gamma=1$  
(for a recent review see \cite{Hands}), 
we conclude that both sets define a chiral conformal field theory in
$d$ and $(D=d+1)$ dimensions respectively. We should also note that
the exponents (\ref{exponents}) are the same as for a zero temperature
dimensionally reduced Gross-Neveu model in which the chiral symmetry
restoration transition will occur as we approach a thermally renormalized
coupling constant obtained during the reduction procedure.

It should be noted that the real solution $\theta=\pi$ makes the model 
to behave as a bosonic model. Thus, this solution transmutate the 
statistics of the fermions into bosons. Unlike to statistical transmutation 
in Chern-Simons models, the present situation is not confined to 
$d=3$.


\section{Conclusions}

We have shown that, even in the absence of canonical scalar fields, we do
obtain non mean field critical exponents for the finite temperature chiral
restoration transition in a four-fermion theory provided we introduce a 
global constraint fixing the particle number. At the saddle-point level, 
it amounts to introducing an imaginary chemical potential which plays the
role of a "gauge" field. This procedure gives rise to a  
zero frequency Matsubara mode which is interpreted, after the limit of high 
temperatures, as the only relevant degree of freedom of the reduced 
theory. The reduced theory is still a four-fermi theory at zero temperature and 
in a lower dimension whose parameters carry a temperature dependence 
resulting from the reduction procedure \cite{Landsman}.
It also exhibits dynamical breaking of discrete chiral
symmetry which is restored near the critical thermally renormalized coupling 
constant, in such a way that the universality class is the same as the original 
finite temperature model.

This behavior should also be expected in any model with dynamical symmetry
breaking, such as QCD, and at finite density, since, on general grounds,
the (bulk) behavior of the system at real (fixed) chemical potential $\mu$
is identical to that at a fixed fermion number $B$, provided $B$ is the
mean fermion number for the system at chemical potential $\mu$.


\section{Acknowledgment}

M.B. Silva Neto is grateful to K. Rajagopal for helpful discussions
and to R. Pisarski for useful comments. 
The authors are especially grateful to H. R. Christiansen for reading and commenting the manuscript. This paper was suported by Conselho Nacional de 
Desenvolvimento Cient\'{\i}fico e Tecnol\'ogico do Brasil (CNPq) and 
Funda\c c\~ao Coordenadoria de Aperfei{\c c}oamento de Pessoal de 
N\'{\i}vel Superior (CAPES).



\begin{thebibliography}{60}
\bibitem{discussion} 
For the sake of comparison, it is worth to mention that 
the $\rho=0$ limit of this sum is 
well known in non-relativistic many body theory. In fact, it is the same as 
$-i$ times the particle density per spin for the atomic limit of the Hubbard 
model at half-filling, which is exactly soluble  (J. Hubbard, Proc. Roy. Soc. (London) {\bf A276}, 238 (1963)). We get  
the same result if we forget the integration over momenta, that is, 
by assuming ${\bf q}^2+\Sigma^2=U^2/4$, where $U$ is the Hubbard interaction). 
This gives exactly $-i/2$ which means that for $\rho=0$, 
$b=2\int^{\Lambda}\frac{d^{d}{\bf q}}{(2\pi)^{d}}$.
\end{thebibliography}
\end{document}